# The Case of the Disappearing (and Re-Appearing) Particle

Yakir Aharonov[1,2,4], Eliahu Cohen[3,4*], Ariel Landau[1], Avshalom C. Elitzur[4]

[1] School of Physics and Astronomy, Tel Aviv University, Tel-Aviv 6997801, Israel
yakir@post.tau.ac.il , ariellan86@gmail.com
[2] Schmid College of Science, Chapman University, Orange, CA 92866, USA
[3] H.H. Wills Physics Laboratory, University of Bristol, Tyndall Avenue, Bristol, BS8 1TL, U.K
eliahu.cohen@bristol.ac.uk
[4] Iyar, The Israeli Institute for Advanced Research, POB 651, Zichron Ya'akov 3095303, Israel
avshalom@iyar.org.il

*Correspondence to eliahu.cohen@bristol.ac.uk

*A novel prediction is derived by the Two-State-Vector-Formalism (TSVF) for a particle superposed over three boxes. Under appropriate pre- and post-selections, and with tunneling enabled between two of the boxes, it is possible to derive not only one, but three predictions for three different times within the intermediate interval. These predictions are moreover contradictory. The particle (when looked for using a projective measurement) seems to disappear from the first box where it would have been previously found with certainty, appearing instead within the third box, to which no tunneling is possible, and later re-appearing within the second. It turns out that local measurement (i.e. opening one of the boxes) fails to indicate the particle's presence, but subtler measurements performed on the two boxes together reveal the particle's nonlocal modular momentum spatially separated from its mass. Another advance of this setting is that, unlike other predictions of the TSVF that rely on weak and/or counterfactual measurements, the present one uses actual projective measurements. This outcome is then corroborated by adding weak measurements and the Aharonov-Bohm effect. The results strengthen the recently suggested time-symmetric Heisenberg ontology based on nonlocal deterministic operators. They can be also tested using the newly developed quantum router.*

## Introduction

The Two-State-Vector Formalism (TSVF) enables quantum mechanics to reveal hitherto unknown aspects of quantum reality [1,2]. This especially holds for the quantum values that prevail *between* two measurements. In standard quantum theory, where only measurement makes a value valid, such "unmeasured values" seem to be meaningless. Yet they are not: In a pre- and post-selected ensemble, the initial and final boundary conditions should be treated on equal footing, equally affecting the state in between. Under the resulting two-times inference, such states not only become accessible, but further reveal novel and intriguing properties of quantum reality. We therefore describe, in what follows, the pre- and post-selected system using a two-time state [1,2]:

$$_{t_f}\langle\phi|\ |\psi\rangle_{t_i}, \qquad (1)$$

where $|\psi\rangle_{t_i}$ and $_{t_f}\langle\phi|$ are the pre- and post-selected states, measured at $t_i$ and $t_f$, respectively. We then let these states evolve unitarily forward/backward from the moment of pre-/post-selection to any moment $t_i < t < t_f$ for fully determining the properties of the system. The latter is given by the corresponding weak value [3] of the corresponding operator $A$:

$$A_w(t) = \frac{\langle\phi(t)|A|\psi(t)\rangle}{\langle\phi(t)|\psi(t)\rangle}. \qquad (2)$$

Recently, TSVF has also revived Heisenberg's picture of quantum mechanics [4,5], by suggesting a description of quantum systems through a set of deterministic operators, based on modular momentum in a time-symmetric framework. This picture has proved to be comprehensive and illuminating with respect to quantum (dynamic) nonlocality. Within it, we employ the well-known analogy between double potential-wells and spin systems [6]. This approach simplifies some calculations when representing the position within the two wells via the Pauli-Z operator and the tunneling between the wells via the Pauli-X operator. These two, together with the Pauli-Y operator, will be understood hereinafter to be nonlocal operators, sensitive to the relative phase between the wells. This difference between operators like the Pauli-Z, which can be measured when looking into a single well, and the Pauli-Y, which incorporates the information from two wells, will turn out to be useful in our gedankenexperiment.

In this paper we continue the line of investigation [7] of probing the foundations of quantum mechanics through apparent quantum paradoxes. In particular, we pursue recent gedankenexperiments where weak values coincide with eigenvalues of projective operators [8], as well as pre- and post-selected scenarios involving non-trivial dynamics [9]. Specifically, we analyze a simple potential-well system where a unique occurrence is shown to take place: a particle within a box seems to "disappear" at a certain instant – leaving behind only its bare nonlocal properties – and then to "reappear" to in another place and re-assume them. Even within the well-known multitude of earlier quantum paradoxes, this evolution is extremely counterintuitive. It is made possible only due to the novel inclusion of tunneling within the pre- and post-selected system.

This work also continues the conceptual advance made by recent works [8,9] of proving TSVF predictions with ordinary (projective) quantum measurements rather than (or in addition to) weak measurements. The importance of this advance is twofold: Theoretically, it avoids criticism of weak measurements raised so far [10-12]. Experimentally, it has already won impressive realization in the work of [13], hence calling for similar realizations of other TSVF predictions, in particular the present one.

This article's outline is as follows. In the first two sections "The Quantum Three Caskets Riddle" and "A Time-Symmetric Calculation" we present a novel effect predicted by the TSVF. In "What has Happened?" and "The Proof" the effect is analyzed. The section "Alternative Accounts" considers other possible explanations and argues that they are insufficient, while "Measuring the Effect" presents two additional verifications of the effect with the aid of weak measurements and the Aharonov-Bohm effect. "The Heisenberg Ontology" relates the recently suggested time-symmetric Heisenberg ontology with the present case. "Recent Experimental Realization" describes a recent experimental breakthrough, of immediate applicability to the present setting.

## Methods and Results

### The Quantum Three Caskets Riddle

Like Shakespeare's Portia presenting the three Caskets to her suitors, let the proverbial Alice of quantum information exercises pose a similar challenge to her Bob. She presents to him three boxes, among which a single particle is hidden. Earlier she has made one out of three measurements on these boxes, and now she discloses to Bob the outcomes of only the first and last measurements, performed at $t=0$ and $t=t_f$, respectively. She then challenges him to find the outcome of the intermediate measurement, performed in between.

*Measurement A*

The pre-selection measurement (Fig. 1) was actually a preparation, splitting the wavefunction without measurement, that sets the stage for the riddle (For illustration we use a three-port beam splitter similar to those in [14,15]. In practice, it could be simpler to use a nested Mach-Zehnder Interferometer as in [16,17]. In any case, we keep the present experiment at the gedanken level).

Alice has prepared a particle at $t = 0$ in the state

$$|\psi(0)\rangle = \frac{1}{\sqrt{3}}(|1\rangle + i|2\rangle + |3\rangle), \tag{3}$$

and placed it superposed within three boxes. Tunneling was enabled between boxes 1 and 2 (but not with 3), such that the effective dynamics is given by the Hamiltonian

$$H = \varepsilon \sigma_x, \tag{4}$$

allowing a minor flipping rate $\varepsilon$ from spin up to spin down and vice-versa, that is, the particle flips its position within Boxes 1 and 2 every $T = \pi\hbar/2\varepsilon$. In terms of box occupations, the eigenstates of $\sigma_x$ are $(1/\sqrt{2})(|1\rangle \pm |2\rangle)$.

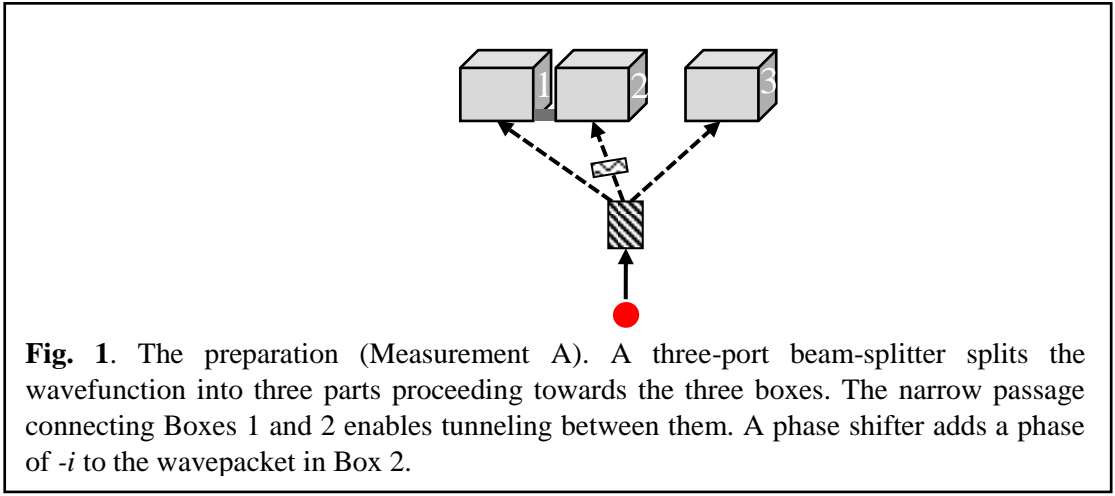

**Fig. 1**. The preparation (Measurement A). A three-port beam-splitter splits the wavefunction into three parts proceeding towards the three boxes. The narrow passage connecting Boxes 1 and 2 enables tunneling between them. A phase shifter adds a phase of -i to the wavepacket in Box 2.

*Measurement B*

This intermediate measurement, which Bob has to derive its outcome from those of A and C, was actually one out of three possible ones, chosen by Alice:

i) If the measurement was performed shortly after A ($t = t_1$ very close to $t = 0$), she has opened only Box 1 (never 2 or 3) to see if the particle is there.

ii) If she has waited till $t = t_2 = \frac{\pi\hbar}{4\varepsilon}$, i.e. half of the tunneling time between boxes 1 and 2 (not to be confused with half of the total experiment time), then she has opened both 1 and 2 (in whatever order) for the same purpose.

iii) If she has further waited until $t = t_3 = \frac{\pi\hbar}{2\varepsilon}$ (when a full flip occurs), she opened only the second box (never 1 or 3).

Only one of these three measurements is allowed for each particle.

*Measurement C*

Finally, Alice has post-selected the particle at $t = t_f = \frac{\pi\hbar}{\varepsilon}$ with the state

$$|\phi(t_f)\rangle = \frac{1}{\sqrt{3}}\left(-|1\rangle + i|2\rangle + |3\rangle\right). \tag{5}$$

This outcome occurs in 1/9 of the cases. Other outcomes were discarded.

*The Challenge*

Being told the outcomes of the initial and final A and C, Bob has to derive *with certainty* those of the three optional intermediate measurements $B_i$, $B_{ii}$ or $B_{iii}$., *i.e.* telling Alice:

"If you chose to make the measurement at $t_1/t_3$, then you have (not) found the particle in Box(es) 1 (and 2). If you performed the measurement at $t_2$ you found again the particle, but in Box 3. And if you have waited further till $t_3$ you found it in Box 2."

**A Time-Symmetric Calculation**

How can Bob find the answer? He has two options, namely to calculate the probabilities in the standard way from past to future, or use conditional probabilities in the time-symmetric way: *Take both measurements, performed before and after the intermediate measurement in question, as equally affecting it* [1]. This contrasts with classical mechanics where initial conditions specify the complete information regarding the system at all later times. It accords, however, with mathematical reasoning, which is indifferent to time's direction (see Fig. 2), and moreover with standard QM where the initial wavefunction does not fully determine the outcomes of all subsequent measurements. This is where TSVF offers its innovation: Add the *future* boundary condition for the complementary information. The resulting account is often surprising, yet in several cases can be corroborated with strong measurements [1,2]. As will be discussed in the next two sections, this allows Bob to find all three outcomes with certainty.

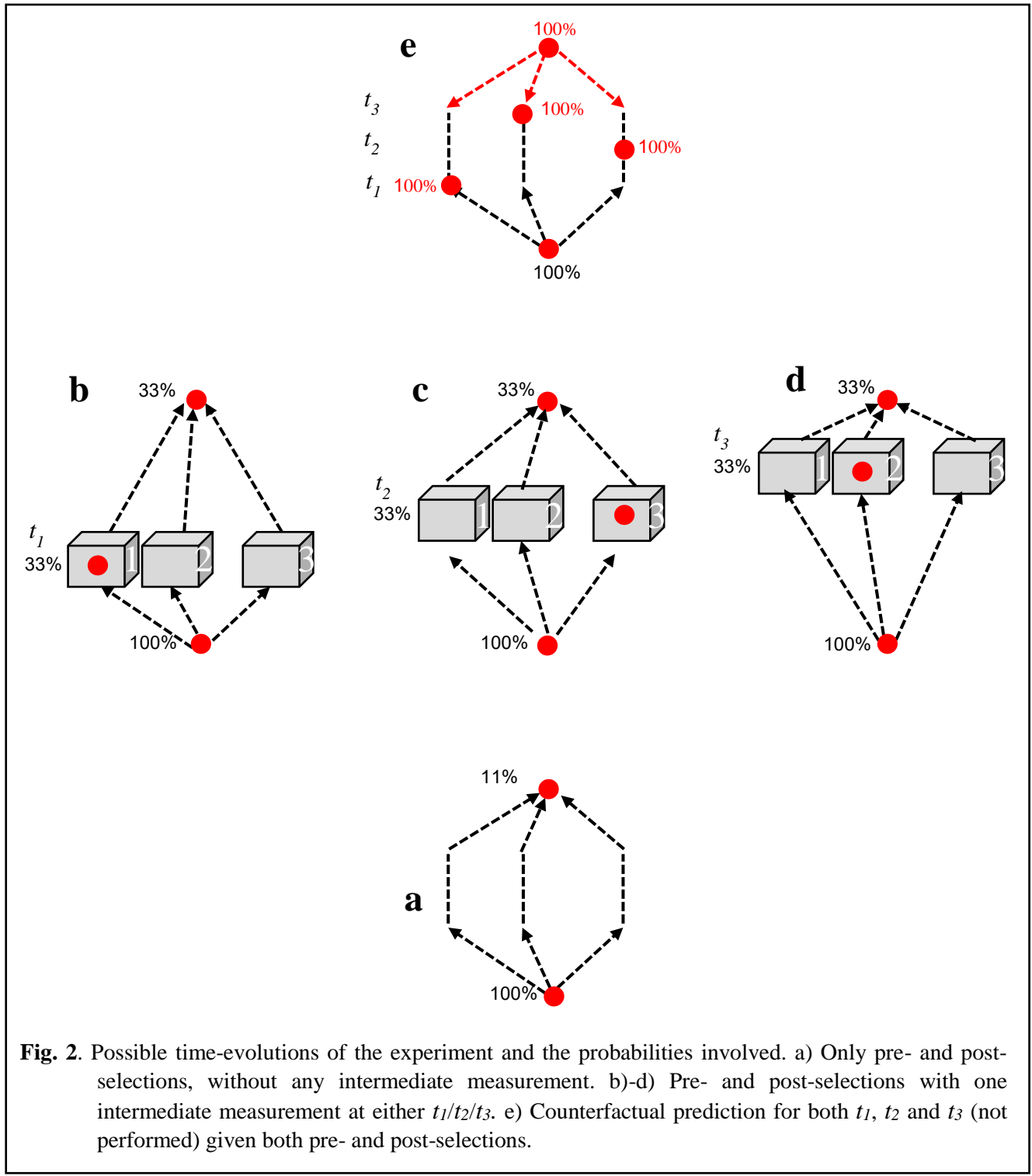

**Fig. 2**. Possible time-evolutions of the experiment and the probabilities involved. a) Only pre- and post-selections, without any intermediate measurement. b)-d) Pre- and post-selections with one intermediate measurement at either $t_1/t_2/t_3$. e) Counterfactual prediction for both $t_1$, $t_2$ and $t_3$ (not performed) given both pre- and post-selections.

**What has Happened?**

Bob's calculation proves correct. Alice confirms that all particles that ended up with the post-selection C shared the following history (Fig. 2): If she chose to make measurement $B_i/B_{iii}$, she always found the particle in Box 1/2, respectively. But if she performed $B_{ii}$, then at $t_2$ she found the particle in neither.

This is obviously an odd evolution. With zero tunneling probability between the first two boxes and the third, the particle could not move from Box 1, where it would have been found with *certainty* at $t = t_1$ nor from Box 2, where it would have been found with certainty at $t = t_3$, to Box 3. As no current is expected between these boxes, no such jump is possible either.

Here then is the reasoning for our predictions (the presentation of the detailed time evolution is postponed until Eqs. 9-10). Let the state of the system be described by a two-time state (see Eq. 1). Let us ignore the normalization from now on, as it does not affect the weak values. The two-time state is evolving due to the tunneling between 1 and 2, such that at $t_1$ it is

$$\langle\phi(t_1)| \ |\psi(t_1)\rangle = \left(\langle 1| + \langle 2|i + \langle 3|\right)\left(|1\rangle + i|2\rangle + |3\rangle\right). \tag{6}$$

Hence at $t_1$ both wavefunctions, from past and future, have a positive amplitude in Box 1 (the weak value of the projection operator is 1), implying the existence of a particle there.

These wavefunctions, however, contain also a unique, *imaginary* amplitude in Box 2 (the weak values of the projection being -1), implying the "negative" existence of the particle there. This element, crucial for understanding this result, is explained in detail below. This also indicates the contextuality embedded in our experiment [18,19].

Overall, a strong measurement of the particle in box 1 at $t_1$ would find it there with certainty, but also, in "weak reality," the total number of particles within the two boxes is zero. This is consistent with the prediction at $t_2$, when the two wavefunctions strongly differ: The two-time state then is

$$\langle\phi(t_2)| \ |\psi(t_2)\rangle = \left(\langle 2|i\sqrt{2} + \langle 3|\right)\left(\sqrt{2}|1\rangle + |3\rangle\right), \tag{7}$$

*i.e.*, the information coming from the past tells us that the particle is in Box 1, while the information coming from the future tells us that it is in Box 2 *with an imaginary amplitude*. This combination of past and future information (in the form of the two-time state, or the individual weak values) tells us (Fig. 3) that the particle, due its tunneling from Box 1 to Box 2, where its existence is "negative," resides in *neither*.

However, if Alice waits until $t_3 = \dfrac{\pi \hbar}{2\varepsilon}$, she will find that, with the wave-function's "positive" and "negative" parts tunneling back between Boxes 1 and 2, the particle now certainly resides in Box 2, obliged by the two-time state

$$\langle \phi(t_3) |\ |\psi(t_3)\rangle = \left( \langle 2|i - \langle 1| + \langle 3| \right)\left( |1\rangle - i|2\rangle + |3\rangle \right). \tag{8}$$

This unique evolution is a direct consequence of the two amplitudes' continuous change in time, due to which, from the very first moment, they were self-cancelling. This self-cancellation, strongly resonating with some previous works such as [20], offers the key for understanding the particle's intriguing evolution. First, this is not an instantaneous "jump" between the boxes. Has the particle been (weakly or strongly) measured in Box 3, it would always be found there. In Boxes 1 and 2, however, the situation is much subtler. *Together*, they contain zero particles. At $t_1$, this zero is composed of +1 in the first box and -1 in the second. Then at $t_2$, due to the tunneling allowed between these boxes, it turns into two zeroes in both, which is what a local strong measurement would reveal.

This transition has evolved continuously in time. The pre-selected wavefunction evolved according to

$$\begin{aligned}|\psi_i(t)\rangle = e^{-iHt/\hbar}|\psi_i(0)\rangle &= \begin{pmatrix} \cos(\varepsilon t/\hbar) & -i\sin(\varepsilon t/\hbar) & 0 \\ -i\sin(\varepsilon t/\hbar) & \cos(\varepsilon t/\hbar) & 0 \\ 0 & 0 & 1 \end{pmatrix}\begin{pmatrix} 1 \\ i \\ 1 \end{pmatrix} = \\ &= \begin{pmatrix} \cos(\varepsilon t/\hbar) + \sin(\varepsilon t/\hbar) \\ i\cos(\varepsilon t/\hbar) - i\sin(\varepsilon t/\hbar) \\ 1 \end{pmatrix},\end{aligned} \tag{9}$$

and similarly for the post-selected wavefunction:

$$\langle \phi_f(t')| = \left( -\cos(\varepsilon t'/\hbar) - \sin(\varepsilon t'/\hbar) \quad -i\cos(\varepsilon t'/\hbar) + i\sin(\varepsilon t'/\hbar) \quad 1 \right), \tag{10}$$

where $t'=0$ at the time of post-selection and grows backwards. When reduced to the first two boxes, these wavefunctions remain orthogonal at any instant. Together they suggest the continuous change in the particle number within each of the boxes, with the total remaining 0 at all times.

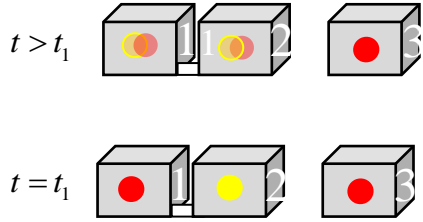

**Fig. 3**. The story told by weak values. At $t = t_1$ the weak values within the three boxes are 1,-1, and 1. Later in time, the weak values within the first two boxes "mix," (*i.e.* the left becomes < 1 and the right becomes > -1, but their sum remains 0) and the weak value in the third box remains 1.

Notice that the "-1 particle" should not be confused with an anti-particle. In contrast to the latter, which differs from its particle only with charge, a "-1 particle" is unique in that all its properties are negative when probed by weak measurement. This is an intriguing prediction of the TSVF, implicit already in the original 3-boxes paradox [21] and in Vaidman's nested Mach-Zehnder experiment [16,17].

As a final note in this section, let us clarify the meaning of the phrase "with certainty" appearing repeatedly within the text. Although less naturally, it can be understood also within a single-state-vector approach. At $t_1$ for instance, we claim that under our pre- and post-selection, the particle will always be found in Box 1, if we look for it there. Why is that? Let us assume, Reductio ad absurdum, that the particle is not there. Hence, we are left with the following contribution for the initial state $(1/\sqrt{2})(i|2\rangle+|3\rangle)$. At the moment $t_f = \pi\hbar/\varepsilon$, this state evolves to $(1/\sqrt{2})(-i|2\rangle+|3\rangle)$, which is orthogonal to the post-selected state, hence post-selection never succeeds in such cases. The same can be easily shown for the measurements at $t_2$ and $t_3$. Hence, a single-state-vector approach works just as well, but we find the two-state approach more intuitive and mathematically simpler.

**The Proof**

An important theorem [21] proves the following for weak values revealed by counterfactual strong (projective) measurements: If a strong measurement's outcome is known with certainty, it equals the outcome of a corresponding weak measurement. For a class of dichotomic operators, *i.e.* operators with only two eigenvalues, the inverse is also true [21]: If the weak value coincides with an eigenvalue of the dichotomic operator, it could also be found using a strong (projective) measurement.

This theorem allows discussing the present paradox and its solution in terms of strong measurements first.

In the two-times Heisenberg picture, all measurement outcomes correspond to the pre- and post-selected case (a subscript *w* could have been added to remind that we are calculating weak values, but then again, all results are deterministic and could have been found in the strong sense as well). This picture is especially helpful in the present case, which is based on simple dynamics resembling classical precession.

Boxes 1 and 2 are described using the set of operators $\{I', \sigma'_x, \sigma'_y, \sigma'_z\}$, which in the larger Hilbert space of the three boxes take the form:

$$I' = \begin{pmatrix} 1 & 0 & 0 \\ 0 & 1 & 0 \\ 0 & 0 & 0 \end{pmatrix}, \quad \sigma'_x = \begin{pmatrix} 0 & 1 & 0 \\ 1 & 0 & 0 \\ 0 & 0 & 0 \end{pmatrix}, \quad \sigma'_y = \begin{pmatrix} 0 & -i & 0 \\ i & 0 & 0 \\ 0 & 0 & 0 \end{pmatrix}, \quad \sigma'_z = \begin{pmatrix} 1 & 0 & 0 \\ 0 & -1 & 0 \\ 0 & 0 & 0 \end{pmatrix}. \quad (11)$$

(The last row and column are zero, since the time evolution does not affect the third box).

As noted above, the $\sigma'_z$ operator denotes a *local* position measurement, revealing whether the particle exists in the first/second box upon inspection. The operators $\sigma'_x$ and $\sigma'_y$ are used to perform *nonlocal* measurements of the relative phase (or modular momentum) between the two boxes [4,5].

The particle's presence within the first box can be measured either weakly (see Sec. "Measuring the Effect" below) or directly (*e.g.* via a scattering experiment) at time $t = t_1 \approx 0$ using a projective measurement performed at this moment:

$$\Pi_1(t_1) \equiv |1\rangle\langle 1|(t_1) = \frac{I' + \sigma'_z}{2}(t_1) = \frac{0+2}{2} = 1, \quad (12)$$

indicating with certainty the particle's *presence* (where the operators are evaluated in the two-time sense). Note that, since the pre- and post-selected states are orthogonal within the first two boxes, the reduced identity operator in this sub-space is actually null.

Similarly, if Alice decides to open the second box at $t = t_3 = \dfrac{\pi\hbar}{2\varepsilon}$ she finds it there with certainty:

$$\Pi_2(t_3) \equiv |2\rangle\langle 2|(t_3) = \frac{I'+\sigma'_z}{2}(t_3) = \frac{0+2}{2} = 1. \tag{13}$$

If, alternatively, the particle is looked for at $t = t_2 = \frac{\pi\hbar}{4\varepsilon}$ in either or both 1 and 2 using a "strong" projective measurement (where again the "identity" operator is equal to zero within the first and second boxes), then

$$\Pi_1(t_2) \equiv |1\rangle\langle 1|(t_2) = \frac{I'+\sigma'_z}{2}(t_2) = \frac{0+0}{2} = 0, \tag{14}$$

$$\Pi_2(t_2) \equiv |2\rangle\langle 2|(t_2) = \frac{I'-\sigma'_z}{2}(t_2) = \frac{0+0}{2} = 0, \tag{15}$$

indicating now its *absence* from both boxes.

Despite its absence, however, the particle has left a trace within these two boxes. An indication of its subtle presence in them is given by the nonlocal modular momentum operator

$$\sigma'_y(t_2) = -2, \tag{16}$$

making the effect purely quantum. This single-particle nonlocal property has no classical analogue [4,5]. Its inference with the aid of weak measurements is discussed in Sec. "Measuring the effect" below.

Upon local projective measurement, only the local properties of the particle, described by $\sigma'_z$ can be found in Box 3. Yet its nonlocal properties, described by $\sigma'_x$ and $\sigma'_y$, reside in 1 and 2, manifesting an odd separation between the particle and its modular momentum. This resembles the quantum Cheshire cat [22], although in our case only the mechanical properties of the particle are discussed, without referring to an inner degree of freedom such as spin.

We conclude this section with a note. If Bob delays the post-selection until $t_f' = (\pi + 2\pi k)\varepsilon/\hbar$, where $k$ is some integer, Alice has yet another option: Instead of one triplet of times $t_1$, $t_2$, $t_3$ she can have an *endless series* of such triplets out of which she can choose any $t_i' = t_i + 2\pi k_i \varepsilon/\hbar$ for $i=1,2,3$, to perform any measurement $B_i$ as with the first three. All the above predictions hold for the new triplet as well.

**Alternative Accounts**

Can, then, this contradictory triplet of outcomes – namely,

i.  If you measure the particle's position at $t_1$ in Box 1 it will always be there;
ii. But if you measure it at $t_2$ in Boxes 1 and 2 it will never be in either;
iii. Yet if you measure it at $t_3$ in Box 2 it will reappear there,

be accounted for in a more trivial way? Let us give this option fair hearing. One may claim that when we looked for the particle at $t = t_2$, it has *all along* been absent in Boxes 1 and 2 and present in 3.

This alternative, however, can be ruled out by Bell's theorem. Arguing that the particle has been in a certain box *all along* is a local hidden-variable account, involving, say, one particle in one box plus a guide-wave split over the three boxes like in Bohmian Mechanics. This account forbids the corpuscle to instantaneously jump from one box to another. It has, however, been ruled out by gedankenexperiments of the kind proposed by Hardy [23]. Consider, *e.g.*, a photon split as in our pre-selection and then sent towards three atoms. Only one atom becomes excited, but for this reason all three become *entangled* [24]. Now just add, as in the EPR case, another measurement that is orthogonal to excited/ground, and the resulting state will prove that *the exciting photon could not have been heading towards one atom all along*. The same holds for our atom: It could not traverse only one of the three paths. The difference between the three different predictions for $t_1$, $t_2$ and $t_3$ thus invokes a genuine disappearance-reappearance cycle.

Finally, recall that in option (*i*), namely opening Box 1 at $t_1$, Alice must refrain from opening Box 2. This strains the local alternative further to the point of absurdity, unless one allows for hidden variables originating backwards in time from the future post-selection [25].

**Measuring the Effect**

The above analysis can be understood either in terms of counterfactual projective measurements with certain outcomes on pre- and post-selected particles, or in terms of actual projective measurements with uncertain outcomes on preselected-only particles (where the post-selected state in Eq. 5, giving rise to all relevant values, is not guaranteed).

In what follows, we augment these arguments with two additional tests. The methods presented below, based on weak measurements, are again actual, thus allowing measuring all observables with only negligible back-action to the quantum system.

*Weak measurements*

The above gedankenexperiment relied on counterfactual projective measurements. *Weak measurements* [5,26], however, allow analyzing this setup without invoking counterfactuals. In fact, this way we can even perform *all* three measurements, at $t = t_1$ as well as at $t = t_2$ and $t = t_3$, without collapsing the wavefunction, using the suitable pre- and post-selected ensemble. This follows from the unique ability of weak measurements to obtain information about the system without collapsing its state. The theorem cited in Sec. "The Proof" above guarantees that the weak measurements' outcomes, *i.e.* the inferred weak values, will be equal to the projective measurements' outcomes as described by Eqs. 12-15. In other words, weak measurements will confirm the particle's disappearance and re-appearance along its evolution, based on a very weak von Neumann coupling between the measured particle and a measuring pointer. Indeed, weak measurements have been employed in the past for testing the original 3-boxes paradox [27].

*Inserting a solenoid between the first and second boxes*

Assuming now that the particle is charged, in which case we would use electromagnetic fields instead of beam splitters. We can now test the predictions of the above sections in the following way: Let us insert a moving solenoid for a brief moment slightly before $t = t_1$ (see Fig. 4). This addition is very potent because due to the Aharonov-Bohm effect, the solenoid is known to change only the particle's modular momentum within these two boxes, *i.e.* it is changing a nonlocal property. For concreteness let us assume that the solenoid is chosen in such a way that the Aharonov-Bohm phase it induces when the charged particle is encircling it flips the relative phase between the two boxes in the initial state (see for example [28]), that is $\frac{1}{\sqrt{3}}(|1\rangle + i|2\rangle + |3\rangle) \to \frac{1}{\sqrt{3}}(-|1\rangle + i|2\rangle + |3\rangle)$ (alternatively, the phase shift can be induced by a time-dependent electric potential acting on the first box). This changes the weak value in the second box from -1 to +1, hence we would find the particle (in the strong, counterfactual sense) with certainty in Box 2, had we decided to look for it there.

If, alternatively, the solenoid is inserted slightly before $t=t_2$ flipping the phase at that later stage, a weak measurement at $t=t_2$ would reveal this change in the relative phase, leading to a modification of Eq. 16. Now $\sigma'_y(t_2)$ would equal +2, and similarly for $\sigma'_x(t_2)$ being now $-i$ rather than $i$, even though the particle is apparently absent from the two boxes at $t=t_2$ when local projections on the two boxes are employed. Being outside the spectrum of the Pauli-Y matrix, this value of $-i$ can be measured only weakly when repeating the measurement over a large ensemble.

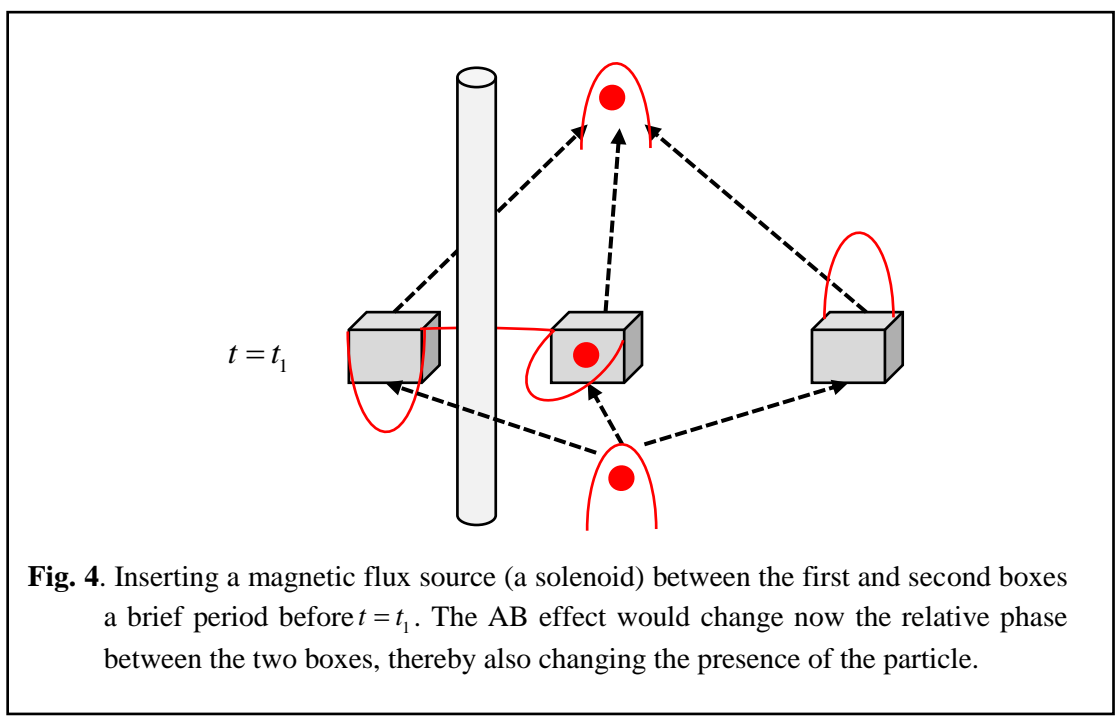

**Fig. 4**. Inserting a magnetic flux source (a solenoid) between the first and second boxes a brief period before $t=t_1$. The AB effect would change now the relative phase between the two boxes, thereby also changing the presence of the particle.

### The Heisenberg Ontology

We believe that the paradox presented here can be best understood within the recently proposed time-symmetric Heisenberg framework [4,5]. Our gedankenexperiment suggests a continuous transition between an ontology based on local properties (such as location in some box) to an ontology based on nonlocal variables (such as modular momentum), which can be verified only by opening two boxes. The latter ontology consists of a set of deterministic operators, that is, a set of operators which would yield with certainty a specific result when measured (this stands in contrast, of course, with an ontology based on the wavefunction). At $t_1$ and $t_3$ the ontological description of the particle is given by the local deterministic operator $\sigma'_z$, while at $t_2$ it is given by the nonlocal deterministic operators $\sigma'_x$ and $\sigma'_y$. On other instances the ontology is given by some combination of local and nonlocal deterministic operators.

### Recent Experimental Realization

As pointed out in the Introduction, the present setting is among the few recent gedanken tests of TSVF's predictions with the aid of the standard, projective ("strong") measurements rather than the customary weak ones. Recently, this theoretical advance has been rewarded with a pioneering experimental realization [13], in a setting resembling the present one, namely, the three boxes paradox [21]. Okamoto and Takeuchi have realized a protocol proposed earlier by Aharonov and Vaidman (AV) for testing the 3-boxes paradox [29]. They managed to turn a photon into a quantum router [30], which can divert another photon similarly to AV's "shutter" designed to reflect it, such that they could demonstrate the photon being reflected from both its possible interaction sites with the particle, fully vindicating AV's prediction.

As the present, "disappearing and reappearing particle paradox" is a variation of the 3-boxes one, with the novel addition of non-trivial time-evolution, it poses a unique challenge for an appropriate adaptation of Okamoto and Takeuchi's experiment. Whereas their setting requires the probing photon to be in spatial superposition, so as to meet the particle in all its "boxes," the present experiment requires a more complex superposition, in both space and time. Let the photon be superposed, first, with respect to its emission time, such that it may be emitted at either $t_1$ towards either Box 1 or Box 3, or at $t_2$ towards Box 3 alone, or at $t_3$ towards either 2 or 3. The parts emitted at $t_1/t_3$ are thus superposed also spatially, splitting the wave-function into five parts in space and time.

The photon is then expected to be reflected from all these varying positions of the tested "shutter" particle. An appropriate re-uniting of the split wave-function in both space and time is therefore expected to prove, by an interference-like revival of the photon's initial state, that it has encountered the particle, indeed, in different places at different times.

## Discussion

We have demonstrated a thought experiment where a particle seems to disappear from a double potential well, despite the zero probability for tunneling outside. Even for those accustomed to quantum oddities, such an effect seems to be order of magnitude weirder.

When analyzed with a time-symmetric formulation of quantum mechanics, the TSVF, this particle is understood to reside all the time within a third potential well, yet with a unique dynamics occurring in the other two. This dynamics suggests that at the beginning, the particle had two, mutually-cancelling amplitudes in the first two boxes, which have later indeed cancelled to zero, hence its disappearance as deduced by local projective measurements. Indeed, simpler and more common phenomena like the Quantum Oblivion effect [20,31], lend further support for this interplay of "events" and "unevents" as an underlying mechanism of many peculiarities of the quantum realm.

An analysis within the time-symmetric Heisenberg picture [4,5], stresses this oddity while suggesting that the particle, although never passing through these wells, has nevertheless left a trace, namely its nonlocal modular momentum. This prediction holds for several complementary methods, including weak and strong (projective) measurements.

The main novelty of this setup is the interplay between local ontology based on local projective measurements of a single location, as in Box 1 at $t_1$, and the nonlocal ontology based on measurements of relative phase between multiple locations, such as Boxes 1 and 2 at $t_2$. We have shown that there is an independent existence of the latter, even in the absence of the former, and indeed, within the recently discussed time-symmetric Heisenberg picture [4,5], both local and nonlocal properties have the same foundational status.

Eventually, we have to accept the existence of both local and nonlocal observables in quantum theory, which makes it inherently distinct from classical mechanics, being only described by local variables. These results accord with other recent findings [3,7,8,9,16,20,21,22,29,31,32] obtained by complementary methods for unique quantum evolutions that seriously undermine the classical nature of time itself.

**Acknowledgements**

It is a pleasure to thank Sandu Popescu and Daniel Rohrlich for helpful comments and discussions. We also wish to thank two anonymous referees for highly constructive remarks. This research was supported in part by Perimeter Institute for Theoretical Physics. Research at Perimeter Institute is supported by the Government of Canada through the Department of Innovation, Science and Economic Development and by the Province of Ontario through the Ministry of Research and Innovation. A.C.E. wishes to especially thank Perimeter Institute's administrative and bistro staff for their precious help. Y.A. acknowledges support from the Israel Science Foundation Grant No. 1311/14 and ICORE Excellence Center "Circle of Light". E.C. was supported by ERC-AdG NLST.

**Author contributions**

Y.A. conceived the main idea and the intuition behind it. A.C.E, E.C. and A.L. further developed it in various aspects. E.C. performed the calculations. A.C.E and E.C. were the main writers, but all authors contributed to the writing and the presentation of the work.

**Competing financial interests**

The authors declare no competing financial interests.